# Knowledge Management in Management of Social and Economic Development of Municipalities: Highlights

Maria A. Shishanina[1], Anatoly A. Sidorov[1]
[1]Tomsk State University of Control Systems and Radioelectronics
Tomsk, Russian Federation
mariia.a.shishanina@tusur.ru

*Abstract* — The paper discusses the process of social and economic development of municipalities. A conclusion is made that developing an adequate model of social and economic development using conventional approaches presents a considerable challenge. It is proposed to use semantic modeling to represent the social and economic development of municipalities, and cognitive mapping to identify the set of connections that occur among indicators and that have a direct impact on social and economic development.

## I. Introduction

The territorial reality of the Russian Federation is such that management of social and economic development (SED) of regions and individual municipalities cannot rely on solutions that are considered standard in the world practice. The unique aspects of development of municipalities determine an essential role of regional and municipal authorities in addressing the problems of social and economic development of territories. However, the problem is exacerbated by the fact that, after the market transformations that the country has experienced, state and municipal authorities were not fully prepared to build an effective region-municipality relationship, which is explained by various regulatory and legal circumstances, financing and the lack of a comprehensive system of information and analytical support that would make it possible to forecast development of territories. In that context, scientific justification of decision-making in municipal SED is a high-priority objective regardless of the development level of any specific territories.

## II. Application of Semantic and Cognitive Tools in the Management of Social and Economic Development of Municipalities

Analysis of the municipal SED process shows that the management model consists of multiple levels. At the same time, [1] notes that semantic models represent the best solution for modeling of SED, as they offer a visual representation that is closest to the natural language.

Based on the semantic network shown generally in figure 4, it should be noted that the SED strategy of a municipality depends on its type, its current level of SED, and on the number of rural settlements included in the municipality.

Analysis of research with a focus on SED shows that any patterns are attributable to geographic and legal aspects of development of the Russian Federation. Based on this conclusion, the type of a municipality is described by a number of factors that can be formally represented as a set $<K, P, A>$ where $K$ is a set of climate zones; $P$ is a set of municipality classifications based on their population; $A$ is a set of municipality classifications based on their specialization. Thus, a specific type of municipality can be described as follows: $T_{kpa} = K \cap P \cap A$.

The current level of SED of a municipality is assessed using a number of indicators. Figure 4 shows that it is reasonable to base the assessment on general indicators that are applicable to all municipality types (e.g., demographics), and on special indicators, the combination of which will depend on the specific municipality type (e. g., indicators that reflect the predominant development of forestry or agriculture).

The set of indicators for assessment of municipal SED are interdependent (e.g., demographics influence indicators that describe the development of social infrastructure, etc.), which should be taken into account when developing the model of municipal SED. At the same time, the semantic network does not make it possible to determine the strength of connections between indicators, which means that this type of relations can be described using the tools of cognitive modeling.

As noted in [2-4], cognitive modeling of semistructured systems is one of the dimensions of the modern theory of decision support that make it possible to achieve adequate results with a large number of interdependent factors. The process of cognitive modeling has several stages, the main one being to identify the set of factors and the relationship between them in a semistructured system. In this case, SED indicators of a specific territory serve as the set of factors for the cognitive model. A standard cognitive model (figure 5) was built based on a set of indicators used to assess the SED, which model can serve as a basis for modeling of SED development in specific territories.

Based on this model, a conclusion can be made that the quality of life should be used as the target factor, since federal-level strategic documents identify it as a key factor in municipal SED. "Production" can be used as a variable (special) factor, because it will be showing significant variation depending on the specialization of any given territory (e.g., settlements can be specialized in agriculture, mining, and so on). Each of the factors can be broken down, revealing new levels, links and relationships within the model.

In turn, analysis of various types of cognitive maps [5-7] (Table 1) shows that their tool set is quite adaptive.

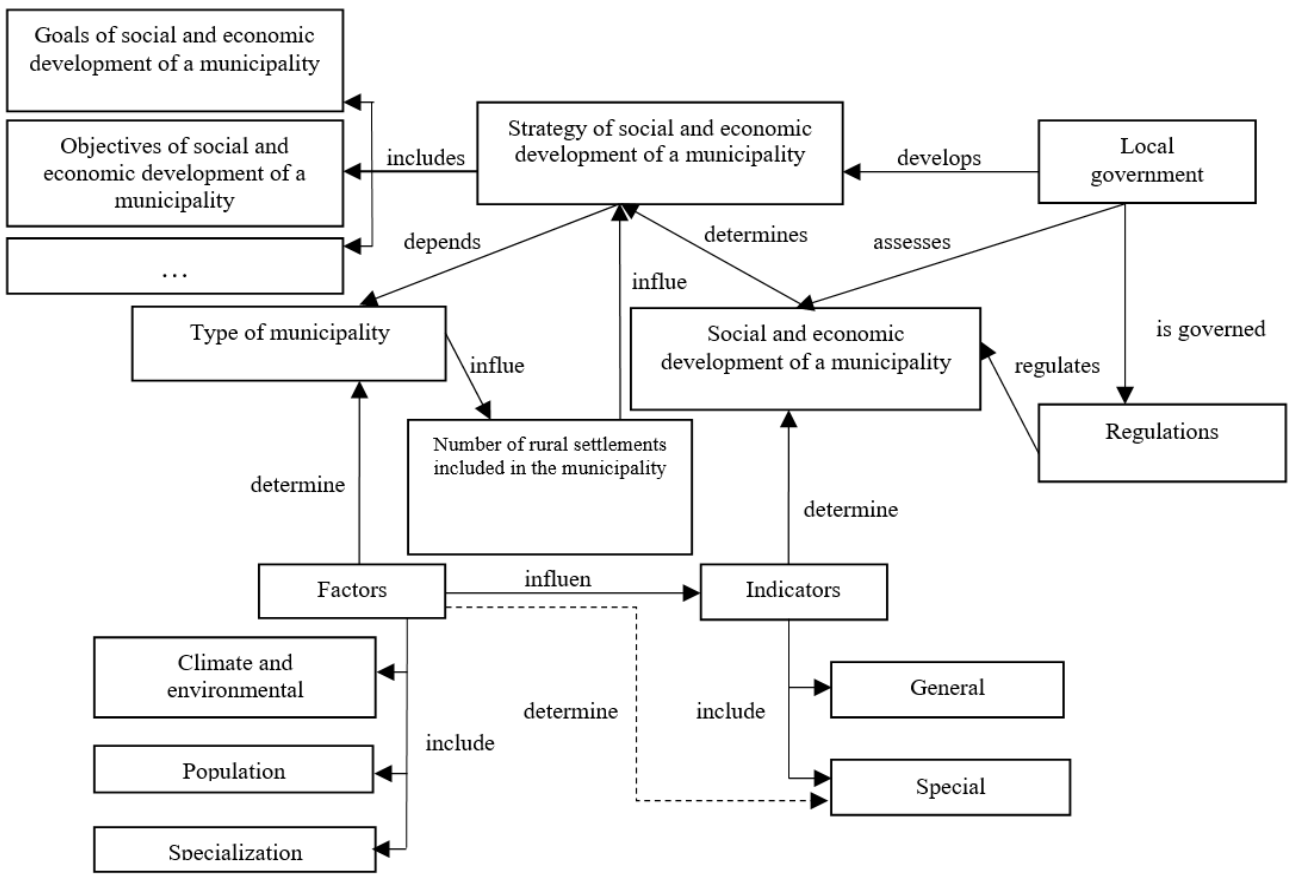

Fig. 4. Semantic network of municipal SED management

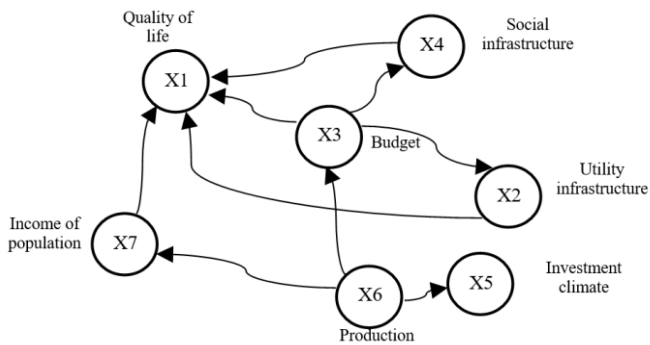

Fig. 5. Standard cognitive model of SED

However, based on these criteria, Silov's fuzzy cognitive maps and fuzzy production cognitive maps are the better choices. The former are a preferred choice for the purposes of modeling, since the latter are rarely used for design and analysis of semistructured systems. In this regard, a fuzzy cognitive map can be represented as follows: $G = <X, W>$, where $X = \{x_1, x_2, \ldots, x_n\}$ corresponds to a set of concepts (factors) in the subject area (in this case, a set of indicators for evaluation of municipal SED), and $W$ is relations over the set $E$ that determine the combination of connections between elements of the given set.

TABLE I. COMPARATIVE ANALYSIS OF TYPES OF COGNITIVE MAPS

| Criterion | Convention al symbolic cognitive maps | Kasko's fuzzy cognitive map | Silov's fuzzy cognitive map | Fuzzy production cognitive map | Fuzzy relational cognitive map |
|---|---|---|---|---|---|
| 1. Option to set the strength of connection | − | + | + | + | + |
| 2. Option to set the negative strength on connection | + | − | + | + | − |
| 3. Usability in problems with quantitative and qualitative factors | − | + | + | + | + |

Elements $e_i$ and $e_j$ are considered connected by the following relation $w(e_i,e_j) \in W \to [-1,1]$ if any change in the value of factor $e_i$ (cause) results in a change in the value of factor $e_j$ (effect). Thus, $e_i$ is deemed to have an effect on $e_j$.

One advantage of utilizing cognitive maps for management of municipal SED is the opportunity to forecast further developments, which can be represented as a system of equations: $O(t+1) = WP(t)$, where $O(t)$ and $O(t+1)$ are vectors of factor value gain at times $t$ and $t+1$ (momentum vectors). In its turn, the situation status at time $t+1$ is determined from the formula $Y(t+1) = Y(t) + O(t+1)$, where $Y(t)$ is the situation status at time $t$. Thus, the cognitive map makes it possible to address two types of problems: static (analysis of the current situation that includes study of the effect that some indicators have on others, study of overall sustainability of the situation and search for structural changes that could produce stable structures) and dynamic (generation and analysis of possible scenarios of development over time).

### III. CONCLUSION

SED of a territory is a complex and continuous process of planning and forecasting that engages authorities of all levels. However, despite the formalized approach to planning at the federal level, municipalities face a number of problems in the process of this work. The greatest challenge in the process of decision-making in SED is the unavailability of complete information about the territory to the individual decision-maker. That is due to the fact that SED is a semistructured subject area, which fact must be taken into account in the process of planning and forecasting. To neutralize the negative impact of problems identified during the study, the authors propose to use semantic and cognitive tools in practical management of municipal SED, which will ultimately improve the quality and level of justification of managerial decisions made by officials.

### ACKNOWLEDGMENT

This paper is designed as part of the state assignment of the Ministry of Science and Higher Education; project FEWM-2020-0036.

### REFERENCES

[1] A.A. Sidorov and M.A. Shishanina, "Semantic network as a tool for determining the management of social and economic development of municipalities," Electronic Devices and Control Systems, vol. 1 no. 2, 2017, pp. 188-192. , in Russian

[2] G.N.Likhosherstova and R.A. Skachkov, "Regional cluster as a vector of socio-economic development of the territory," Nauchnyi Rezultat Journal, vol. 3, 2015, pp. 91–107.

[3] P.P. Groumpos Fuzzy Cognitive Maps: Basic Theories and Their Application to Complex Systems. In: Glykas M. (eds) Fuzzy Cognitive Maps. Studies in Fuzziness and Soft Computing, Springer, Berlin, Heidelberg, vol 247, 2010, pp.1-23.

[4] S.A. Gray, E. Zanre, and S. R. J. Gray, "Fuzzy cognitive maps as representations of mental models and group beliefs: theoretical and technical issues," in Fuzzy Cognitive Maps for Applied Sciences and Engineering—From Fundamentals to Extensions and Learning Algorithms. Papageorgiou, E. I. ed, 2014, pp. 29–48.

[5] R. Axelrod, Structure of Decision: the cognitive maps of political elites, N.Y., Prinston Univ. Press, 1976.

[6] B. Kosko, "Fuzzy Cognitive Maps", International Journal of Man-Machine Studies, vol. 1, 1986, pp. 65–75.

[7] J. Aguilar "A Dynamic Fuzzy-Cognitive-Map Approach Based on Random Neural Networks," International Journal of Computational Cognition, vol. 1 part. 4, 2002, pp. 48–55.